\documentclass[10pt,english,conference]{IEEEtran}
\usepackage[T1]{fontenc}
\usepackage[latin9]{inputenc}
\usepackage{amsmath}
\usepackage{amsthm}
\usepackage{amssymb}
\usepackage{graphicx}

\makeatletter
\def\ps@IEEEtitlepagestyle{%
  \def\@oddfoot{\mycopyrightnotice}%
  \def\@evenfoot{}%
}
\def\mycopyrightnotice{%
  {\footnotesize 978-1-5090-3009-5/17/$\$31.00$ $©$2017 IEEE \hfill}
  \gdef\mycopyrightnotice{}
}

\theoremstyle{plain}

\theoremstyle{plain}

\ifCLASSINFOpdf
\else
\fi
\usepackage{babel}

\makeatother

\usepackage{babel}
\providecommand{\propositionname}{Proposition}
\providecommand{\theoremname}{Theorem}

\begin{document}

\title{Coded Multicast Fronthauling and Edge Caching for Multi-Connectivity
Transmission in\\ Fog Radio Access Networks}

\author{\IEEEauthorblockN{$^{1}$Seok-Hwan Park, $^{2}$Osvaldo Simeone, $^{3}$Wonju Lee, and
$^{4}$Shlomo Shamai (Shitz)} \IEEEauthorblockA{$^{1}$Division of Electronic Engineering, Chonbuk National University,
Jeonju-si, Jeollabuk-do, 54896 Korea\\
$^{2}$CWiP, New Jersey Institute of Technology, 07102 Newark, New
Jersey, USA\\
$^{3}$Samsung Advanced Institute of Technology, Yongin-si, Gyeonggi-do,
17113 Korea\\
$^{4}$Department of Electrical Engineering, Technion, Haifa, 32000,
Israel\\
 Email: seokhwan@jbnu.ac.kr, osvaldo.simeone@njit.edu, wonjulee@kaist.ac.kr,
sshlomo@ee.technion.ac.il}}
\maketitle
\begin{abstract}
This work studies the advantages of coded multicasting for the downlink
of a Fog Radio Access Network (F-RAN) system equipped with a multicast
fronthaul link. In this system, a control unit (CU) in the baseband
processing unit (BBU) pool is connected to distributed edge nodes
(ENs) through a multicast fronthaul link of finite capacity, and the
ENs have baseband processing and caching capabilities. Each user equipment
(UE) requests a file in a content library which is available at the
CU, and the requested files are served by the closest ENs based on
the cached contents and on the information received on the multicast
fronthaul link. The performance of coded multicast fronthauling is
investigated in terms of the delivery latency of the requested contents
under the assumption of pipelined transmission on the fronthaul and
edge links and of single-user encoding and decoding strategies based
on the hard transfer of files on the fronthaul links. Extensive numerical
results are provided to validate the advantages of the coded multicasting
scheme compared to uncoded unicast and multicast strategies.\end{abstract}

\begin{IEEEkeywords}
C-RAN, F-RAN, edge caching, coded multicasting, latency, multi-connectivity.
\end{IEEEkeywords}

\theoremstyle{theorem}
\newtheorem{theorem}{Theorem}
\theoremstyle{proposition}
\newtheorem{proposition}{Proposition}
\theoremstyle{lemma}
\newtheorem{lemma}{Lemma}
\theoremstyle{corollary}
\newtheorem{corollary}{Corollary}
\theoremstyle{definition}
\newtheorem{definition}{Definition}
\theoremstyle{remark}
\newtheorem{remark}{Remark}

\section{Introduction\label{sec:Introduction}}

\let\thefootnote\relax\footnotetext{S.-H. Park was supported by the NRF Korea funded by the Ministry of Science, ICT $\&$ Future Planning through grant 2015R1C1A1A01051825. The work of O. Simeone was partially supported by the U.S. NSF through grant 1525629. The work of S. Shamai has been supported by the European Union's Horizon 2020 Research And Innovation Programme, grant agreement no. 694630.}

Fog Radio Access Network (F-RAN) is a wireless cellular system that
enables content delivery to user equipments (UEs) by means of both
edge caching and cloud processing \cite{Bi-et-al}-\cite{MPeng-et-al}.
In an F-RAN, edge nodes (ENs), such as small-cell base stations (SBSs),
can pre-fetch frequently requested, or popular, contents for storage
in their local caches, while retrieving uncached information from
the cloud. Prior works \cite{Sengupta-et-al}-\cite{Azimi-et-al}
studied the design of F-RANs from an information-theoretic viewpoint.
Instead, the references \cite{Park-et-al:FRAN-SPAWC}-\cite{Park-et-al:FRAN}
took a signal processing perspectives by focusing on the design of
beamforming strategies under different criteria, such as the delivery
latency \cite{Park-et-al:FRAN-SPAWC}, the network cost \cite{Tao-et-al}
and the delivery rate \cite{Park-et-al:FRAN}. All these papers assumed
that every EN has a dedicated orthogonal fronthaul link to a control
unit (CU) in the cloud.

\begin{figure}
\centering\includegraphics[width=6.35cm,height=6.45cm]{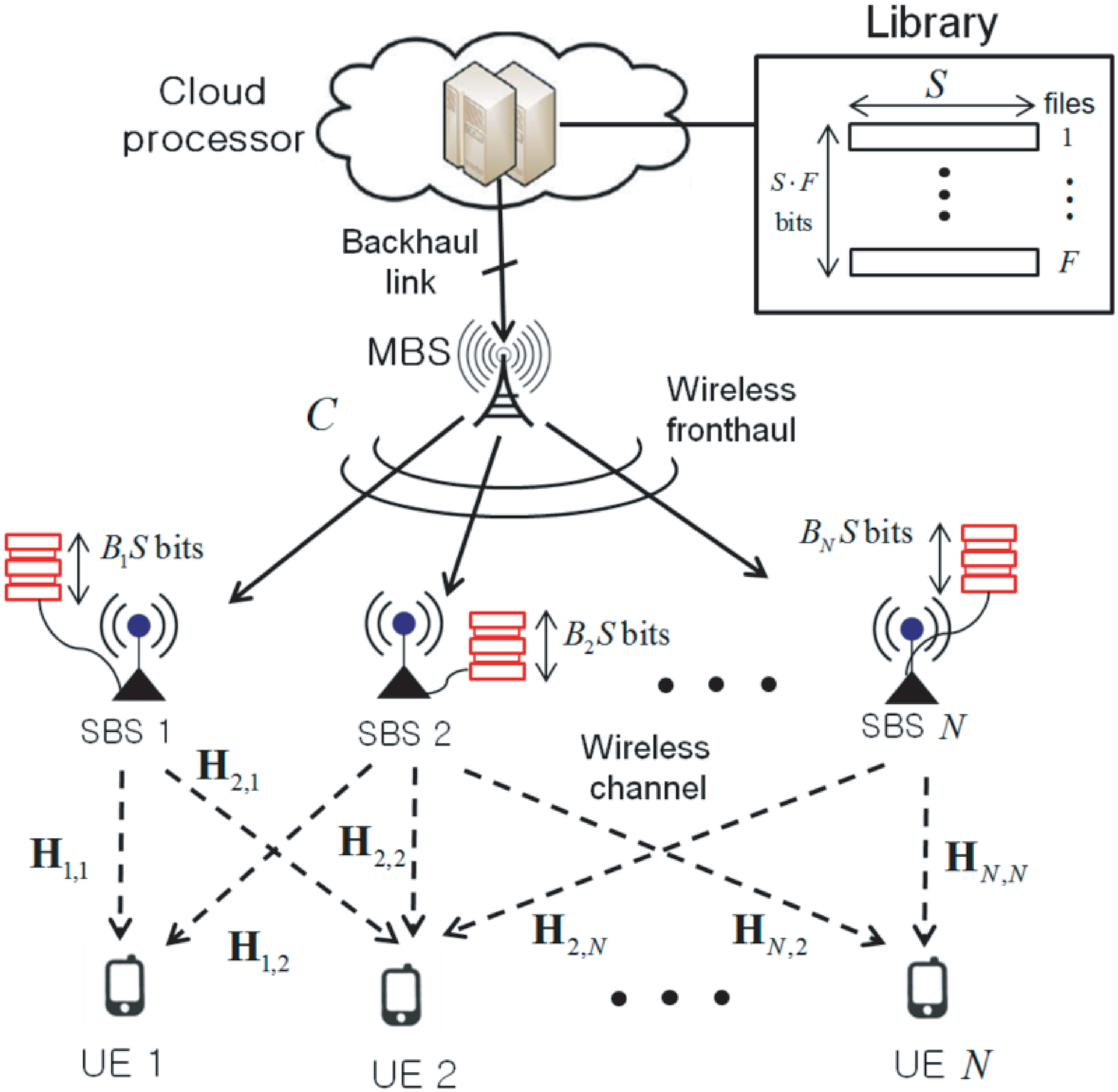}

\caption{\label{fig:System-Model-wireless}{\scriptsize{}Illustration of an
F-RAN model with wireless fronthaul link in which a cloud processor
is connected to small-cell BSs, each equipped with local cache, through
a macro BS which communicates with the small-cell BSs on a wireless
fronthaul link.}}
\end{figure}

\begin{figure}
\centering\includegraphics[width=6.35cm,height=6.45cm]{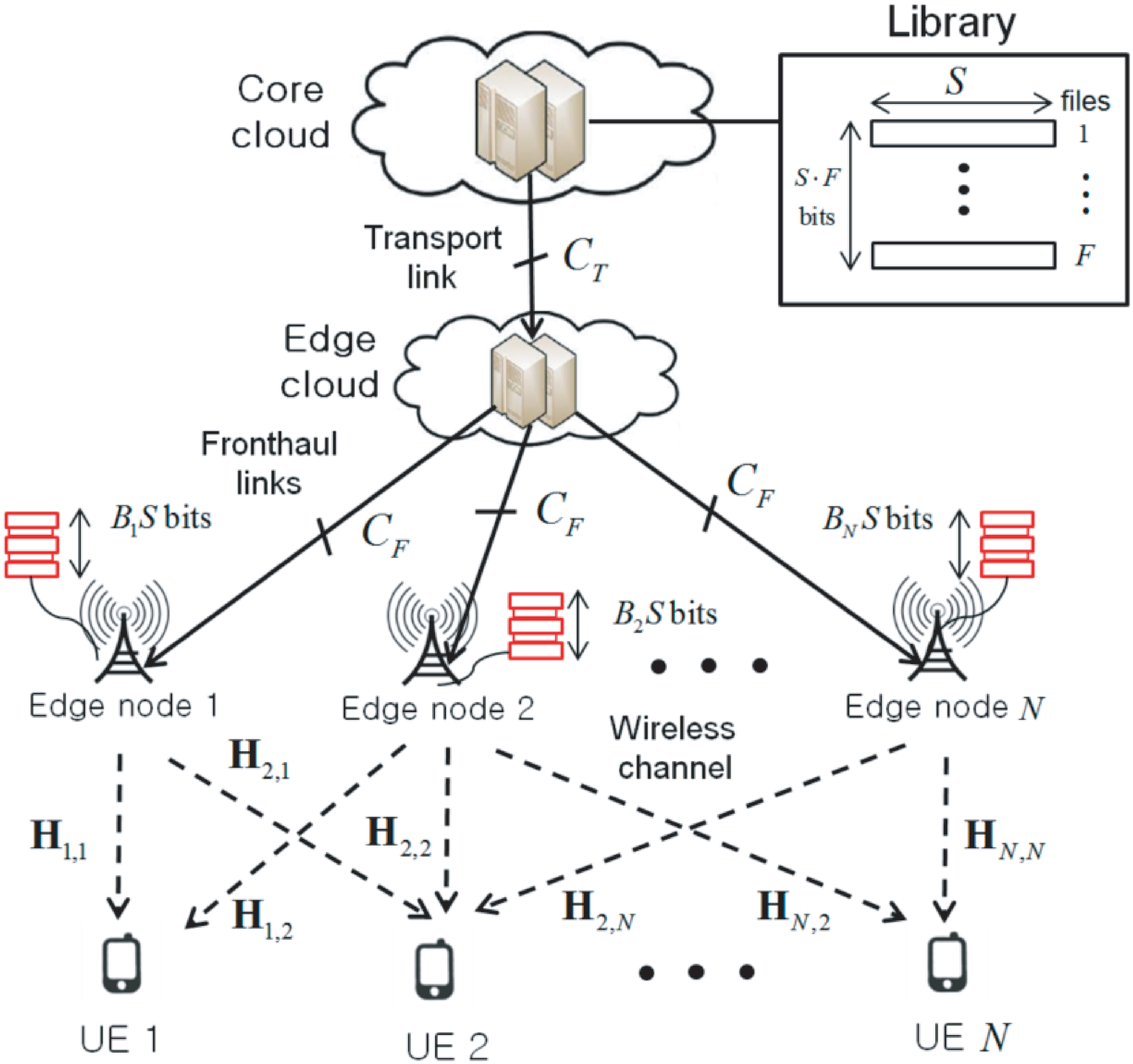}

\caption{\label{fig:System-Model-hierarchical}{\scriptsize{}Illustration of
a hierarchical F-RAN model, in which a core cloud has access to the
content library and is connected to edge nodes, each equipped with
local cache, through an edge cloud unit.}}
\end{figure}

In contrast, in this work, we study an F-RAN system characterized
by a shared multicast fronthaul link from the CU to the ENs. This
model is motivated by two important deployment scenarios, which are
shown in Fig. \ref{fig:System-Model-wireless} and Fig. \ref{fig:System-Model-hierarchical}.
In Fig. \ref{fig:System-Model-wireless}, the cloud processor is connected
via a high-speed backhaul link to a macro BS (MBS) that provides a
wireless fronthaul connection to a number of SBSs \cite{Koh-et-al}.
In the scenario of Fig. \ref{fig:System-Model-hierarchical}, the
cloud has a hierarchical structure with an \textit{edge cloud} (EC)
being connected to a \textit{core cloud} (CC). Since the CC is connected
to the ENs only through the EC, assuming that the latter only forwards
the signal received from the CC, the link from the CC to the ENs can
be equivalently modeled by a multicast link of capacity $C=\min\{C_{T},C_{F}\}$,
where $C_{T}$ and $C_{F}$ are the capacities of the CC-to-EC transport
link and of the EC-to-ENs fronthaul links, respectively.

This work studies the latency needed for the delivery of the files
requested by UEs for an F-RAN with a shared multicast fronthaul link.
In particular, we investigate the advantages of coded multicasting
\cite{Ji-et-al} for fronthaul transmission by focusing on single-user
encoding and decoding strategies based on the hard transfer of files
on the fronthaul links \cite{Park-et-al:FRAN}. With regards to communication
from the ENs to the UEs on the edge wireless channels, we consider
general multi-connectivity strategies as determined by the cached
contents and focus on the design of linear beamforming strategies.
We derive an efficient optimization solution based on the concave
convex procedure (CCCP) and provide extensive numerical results that
validate the advantages of coded multicasting as compared to conventional
uncoded unicasting and multicasting approaches.

\textit{Notation}: We denote the mutual information between the random
variables $X$ and $Y$ as $I(X;Y)$, and the circularly symmetric
complex Gaussian distribution with mean $\mbox{\boldmath${\mu}$}$
and covariance matrix $\mathbf{R}$ as $\mathcal{CN}(\mbox{\boldmath${\mu}$},\bold{R})$.
$\mathbb{C}^{M\times N}$ denotes the set of all $M\times N$ complex
matrices and $\mathbb{E}(\cdot)$ represents the expectation operator.
The operations $(\cdot)^{T}$ and $(\cdot)^{\dagger}$ denote the
transpose and Hermitian transpose of a matrix, and $\left\lfloor x\right\rfloor $
is the largest integer not larger than $x$. The determinant and trace
of a matrix $\mathbf{X}$ are denoted as $\det(\mathbf{X})$ and $\mathrm{tr}(\mathbf{X})$,
respectively.

\section{System Model\label{sec:System-Model}}

As seen in Fig. \ref{fig:System-Model-wireless}, we consider the
downlink of an F-RAN with $N$ pairs of ENs and UEs, where the ENs
are connected to a CU by means of a shared multicast fronthaul link
of capacity $C$ bit/symbol. Henceforth, a symbol refers to a channel
use of the downlink wireless channel. In this system, the UEs request
files from a static library of $F$ popular files, where each file
$f\in\mathcal{F}\triangleq\{1,\ldots,F\}$ is of size $S$ bits. It
is assumed that the CU has access to the library and that the probability
$p(f)$ of a file $f$ to be selected is given by Zipf's distribution
$p(f)=cf^{-\gamma}$ for $f\in\mathcal{F}$, where $\gamma\geq0$
is a given popularity exponent and $c\geq0$ is set such that $\sum_{f\in\mathcal{F}}p(f)=1$.
The file requested by the $k$th UE is denoted by $f_{k}\in\mathcal{F}$,
which is assumed to be independent across the index $k$. We define
the demand vector $\mathbf{f}=[f_{1}\,f_{2}\,\ldots\,f_{N}]^{T}$
and the set $\mathcal{F}_{\mathrm{req}}=\cup_{k\in\mathcal{N}}\{f_{k}\}$
of requested files.

Each EN can cache $B_{i}S$ bits from the library, and we define the
fractional caching capacity $\mu_{i}$ of EN $i$ as
\begin{equation}
\mu_{i}=\frac{B_{i}}{F}.\label{eq:fractional-capacity}
\end{equation}
We focus on the symmetric case of $B_{i}=B$ and $\mu_{i}=\mu$ for
all $i\in\mathcal{N}\triangleq\{1,\ldots,N\}$, but the discussion
can be extended to general cases.

\subsection{Channel model\label{sub:Channel-model}}

We assume that each EN and UE are equipped with $n_{T}$ and $n_{R}$
antennas, respectively. Under a flat-fading channel model, the baseband
signal $\mathbf{y}_{k}\in\mathbb{C}^{n_{R}\times1}$ received by UE
$k$ in each transmission interval is given as
\begin{equation}
\mathbf{y}_{k}=\sum_{i\in\mathcal{N}}\mathbf{H}_{k,i}\mathbf{x}_{i}+\mathbf{z}_{k}=\mathbf{H}_{k}\mathbf{x}+\mathbf{z}_{k},\label{eq:received-signal}
\end{equation}
where $\mathbf{x}_{i}\in\mathbb{C}^{n_{T}\times1}$ is the baseband
signal transmitted by EN $i$; $\mathbf{H}_{k,i}\in\mathbb{C}^{n_{R}\times n_{T}}$
denotes the channel response matrix from EN $i$ to UE $k$; $\mathbf{z}_{k}\in\mathbb{C}^{n_{R}\times1}$
is the additive noise distributed as $\mathbf{z}_{k}\sim\mathcal{CN}(\mathbf{0},\mathbf{I})$;
$\mathbf{H}_{k}\triangleq[\mathbf{H}_{k,1}\ldots\mathbf{H}_{k,N}]\in\mathbb{C}^{n_{R}\times Nn_{T}}$
collects the channel matrices $\mathbf{H}_{k,i}$ from all the ENs
to UE $k$; and $\mathbf{x}\triangleq[\mathbf{x}_{1}^{\dagger}\,\ldots\,\mathbf{x}_{N_{R}}^{\dagger}]^{\dagger}\in\mathbb{C}^{Nn_{T}\times1}$
is the signal transmitted by all the ENs. We assume that each EN $i$
is subject to the average transmit power constraint $\mathbb{E}\left\Vert \mathbf{x}_{i}\right\Vert ^{2}\leq P$,
and the signal-to-noise ratio (SNR) of the wireless edge link is defined
as $P$. Furthermore, the channel matrices $\{\mathbf{H}_{k,i}\}_{k,i\in\mathcal{N}}$
are assumed to remain constant during each transmission interval.
In Sec. \ref{sec:Numerical-Results}, we will evaluate the system
performance under the additional assumption that the elements of the
channel matrix $\mathbf{H}_{k,i}$ are independent and identically
distributed (i.i.d.) as $\mathcal{CN}(0,\alpha^{|k-i|})$, where the
parameter $0<\alpha<1$ accounts for the path loss. Note that this
choice reflects the facts that UE $k$ is closest to its serving EN
$k$ and that the distance between UE $k$ and EN $i$ increases with
the difference $|k-i|$.

\subsection{System Operation and Delivery Latency\label{sub:Pre-Fetching-Phase-and}}

The F-RAN system under study operates in two phases, namely pre-fetching
and delivery, as in \cite{Park-et-al:FRAN} and references therein
(see \cite{Azimi-et-al} for an online problem formulation). In the
pre-fetching phase, which takes place offline, say at night, each
EN $i$ populates its cache from the library of $F$ files. In the
following delivery phase, which spans many time slots, for any given
demand vector $\mathbf{f}$, the CU and ENs cooperate to serve the
UEs via the multicast fronthaul and wireless edge links in each time
slot.

For the pre-fetching phase, we consider the randomized fractional
cache distinct strategy studied in \cite[Sec. III-C]{Park-et-al:FRAN},
in which each EN populates its cache in a distributed manner. With
this strategy, each file $f$ is split into $L$ equal-sized subfiles
$(f,1),\ldots,(f,L)$, such that each subfile $(f,l)$ is of size
$\tilde{S}=S/L$ bits. Each EN $i$ stores randomly chosen $\tilde{L}=\left\lfloor \mu L\right\rfloor $
fragments of every file $f\in\mathcal{F}$. We define binary caching
variables $\mathbf{c}\triangleq\{c_{f,l}^{i}\}_{f\in\mathcal{F},l\in\mathcal{L}}$,
with $\mathcal{L}=\{1,\ldots,L\}$, as
\begin{align}
c_{f,l}^{i} & =\begin{cases}
1, & \mathrm{if}\,\,\mathrm{subfile}\,\,(f,l)\,\,\mathrm{is}\,\,\mathrm{cached\,\,by\,\,EN}\,\,i\\
0, & \mathrm{otherwise}
\end{cases},\label{eq:caching-variable-definition}
\end{align}
which must satisfy the cache memory constraint
\begin{align}
\sum_{f\in\mathcal{F}}\sum_{l\in\mathcal{L}}c_{f,l}^{i}\tilde{S} & \leq BS,\label{eq:cache-memory-constraint}
\end{align}
for each EN $i$.

In the delivery phase, the CU and the ENs cooperate to deliver the
requested files $\mathcal{F}_{\mathrm{req}}$ to the UEs. As in \cite{Park-et-al:FRAN-SPAWC},
we aim at designing the delivery strategies for the fronthaul and
wireless edge links with the goal of minimizing the delivery coding
latency. Specifically, we focus on single-user encoding and decoding
strategies based on the hard transfer of files on the fronthaul links.
This excludes interference management techniques such as dirty paper
coding and cloud-based precoding as in cloud radio access network
systems (see, e.g., \cite[Sec. IV]{Park-et-al:FRAN}). We also consider
pipelined transmission on the fronthaul and edge links \cite[Sec. VII]{Sengupta-et-al},
such that the overall delivery coding latency $T_{\mathrm{total}}$
is given as
\begin{equation}
T_{\mathrm{total}}=\max\left\{ T_{F},T_{E}\right\} ,\label{eq:latency-total}
\end{equation}
where $T_{F}$ and $T_{E}$ represent the coding latency required
to communicate on the fronthaul and edge links, respectively. In the
following sections, we describe the latency metrics $T_{E}$ and $T_{F}$
under multi-connectivity transmission and various multicasting strategies.

\section{Multi-Connectivity Wireless Transmission\label{sub:Latency-on-Edge}}

For the efficient management of inter-UE interference signals on the
wireless channel, we consider multi-connectivity transmission across
the ENs such that UE $k$ is served by the closest $M\leq N$ ENs,
i.e.,
\begin{equation}
\mathcal{N}_{\mathrm{EN},k}\!=\!\left\{ k\!-\!\!\left\lfloor \frac{M-1}{2}\right\rfloor \!,k-\!\!\left\lfloor \frac{M-1}{2}\right\rfloor \!+\!1,\ldots,k+\!\left\lfloor \frac{M}{2}\right\rfloor \!\right\} ,\label{eq:serving-ENs}
\end{equation}
where $M$ is referred to as connectivity level (see, e.g., \cite{Maeder-et-al}).
Note that, in order for this approach to be implemented, the ENs in
$\mathcal{N}_{\mathrm{EN},k}$ should have the information of the
content $f_{k}$ requested by UE $k$ either by means of caching or
via the information received on the multicast fronthaul link.

Accordingly, the vector $\mathbf{x}$ transmitted by the ENs can be
written as
\begin{equation}
\mathbf{x}=\sum_{k\in\mathcal{N}}\mathbf{V}_{k}\mathbf{s}_{k},\label{eq:precoding-model}
\end{equation}
where $\mathbf{s}_{k}\in\mathbb{C}^{n_{S}\times1}$ is the baseband
signal vector that encodes the file $f_{k}$ and is distributed as
$\mathbf{s}_{k}\sim\mathcal{CN}(\mathbf{0},\mathbf{I})$; and $\mathbf{V}_{k}=[\mathbf{V}_{k,1}^{\dagger}\,\ldots\,\mathbf{V}_{k,N}^{\dagger}]^{\dagger}\in\mathbb{C}^{Nn_{T}\times n_{S}}$
is the precoding matrix for the signal $\mathbf{s}_{k}$ with the
submatrix $\mathbf{V}_{k,i}\in\mathbb{C}^{n_{T}\times n_{S}}$ corresponding
to EN $i$. Note that $n_{S}\leq\min\{Nn_{T},n_{R}\}$ represents
the number of data streams that encode any file $f_{k}$, and that
the precoding matrices $\mathbf{V}_{k,i}$ are subject to the connectivity
condition
\begin{equation}
\mathrm{tr}\left(\mathbf{V}_{k,i}\mathbf{V}_{k,i}^{\dagger}\right)=0,\,\,\mathrm{for\,\,all}\,\,(k,i)\,\,\mathrm{with}\,\,i\notin\mathcal{N}_{\mathrm{EN},k}.\label{eq:connectivity-constraint}
\end{equation}
This equality imposes that the file $\mathbf{s}_{k}$ cannot be precoded
by EN $i$ unless the EN belongs to the set $\mathcal{N}_{\mathrm{EN},k}$.

Under the precoding model (\ref{eq:precoding-model}) and the assumption
that each UE $k$ decodes the file $f_{k}$ based on the received
signal $\mathbf{y}_{k}$ in (\ref{eq:received-signal}) by treating
interference as noise, the achievable rate $R_{k}$ for UE $k$ is
given as
\begin{align}
 & R_{k}=g_{k}\left(\mathbf{V}\right)\triangleq I\left(\mathbf{s}_{k};\mathbf{y}_{k}\right)\label{eq:rate-UE-k}\\
 & =\phi\left(\mathbf{H}_{k}\mathbf{V}_{k}\mathbf{V}_{k}^{\dagger}\mathbf{H}_{k}^{\dagger},\,\sum_{l\in\mathcal{N}\setminus\{k\}}\!\!\mathbf{H}_{k}\mathbf{V}_{l}\mathbf{V}_{l}^{\dagger}\mathbf{H}_{k}^{\dagger}+\mathbf{I}\right),\!\nonumber
\end{align}
where we defined the notation $\mathbf{V}\triangleq\{\mathbf{V}_{k}\}_{k\in\mathcal{N}}$
and the function $\phi(\mathbf{A},\mathbf{B})\triangleq\log_{2}\det(\mathbf{A}+\mathbf{B})-\log_{2}\det(\mathbf{B})$.
For given delivery rates $\mathbf{R}\triangleq\{R_{k}\}_{k\in\mathcal{N}}$,
the latency $T_{E}$ on the edge link is given as
\begin{equation}
T_{E}=\frac{S}{\min_{k\in\mathcal{N}}R_{k}},\label{eq:edge-latency}
\end{equation}
since $\min_{k\in\mathcal{N}}R_{k}$ is the rate at which all requested
files $\mathcal{F}_{\mathrm{req}}$ can be delivered to the UEs. We
note that increasing the connectivity level $M$ always improves the
rates $R_{k}$ and thus reduces the edge latency $T_{E}$ by relaxing
the constraints (\ref{eq:connectivity-constraint}). However, we will
see in Sec. \ref{sub:Fronthauling} that this does not guarantee improved
total latency $T_{\mathrm{total}}$ due to the increased fronthaul
overhead.

From (\ref{eq:edge-latency}), we can see that the problem of minimizing
the latency on the edge link is equivalent to that of maximizing the
minimum rate $R_{\min}\triangleq\min_{k\in\mathcal{N}}R_{k}$. Thus,
we consider the problem:\begin{subequations}\label{eq:problem-edge}
\begin{align}
\underset{\mathbf{V},R_{\min}}{\mathrm{maximize}} & \,\,\,R_{\min}\label{eq:problem-edge-objective}\\
\mathrm{s.t.}\,\, & R_{\min}\leq g_{k}\left(\mathbf{V}\right),\,\,k\in\mathcal{N},\label{eq:problem-edge-constraint-rate}\\
 & \mathrm{tr}\left(\mathbf{V}_{k,i}\mathbf{V}_{k,i}^{\dagger}\right)\leq0,\,\,k\in\mathcal{N},\,\,i\notin\mathcal{N}_{\mathrm{EN},k},\label{eq:problem-edge-constraint-connectivity}\\
 & \sum_{k\in\mathcal{N}}\mathrm{tr}\left(\mathbf{V}_{k,i}\mathbf{V}_{k,i}^{\dagger}\right)\leq P,\,\,i\in\mathcal{N}.\label{eq:problem-edge-power}
\end{align}
\end{subequations}To tackle the non-convex problem (\ref{eq:problem-edge}),
as in \cite{Park-et-al:FRAN}, we restate the problem with respect
to the variables $\tilde{\mathbf{V}}_{k}=\mathbf{V}_{k}\mathbf{V}_{k}^{\dagger}$
and relax the constraint $\mathrm{rank}(\tilde{\mathbf{V}}_{k})\leq n_{S}$.
Then, we obtain a difference-of-convex problem and hence can derive
an iterative algorithm based on the CCCP approach that gives non-decreasing
objective values with respect to the number of iterations (see, e.g.,
\cite{Tao-et-al}). After convergence, we obtain the precoding matrices
$\mathbf{V}_{k}$ by taking the $n_{S}$ leading eigenvectors of $\tilde{\mathbf{V}}_{k}$
multiplied by the square roots of the corresponding eigenvalues. Details
follow as in \cite{Park-et-al:FRAN}.

\section{Fronthaul Delivery Strategies\label{sub:Fronthauling}}

In this subsection, we discuss fronthauling strategies on the multicast
fronthaul link and derive the corresponding latency metrics. To elaborate,
we define a binary variable $d_{f_{k},l}^{i}$ as
\begin{equation}
d_{f_{k},l}^{i}=\begin{cases}
1, & i\in\mathcal{N}_{\mathrm{EN},k}\,\,\mathrm{and}\:c_{f_{k},l}^{i}=0\\
0, & \mathrm{otherwise}
\end{cases},\label{eq:binary-transfer-variable}
\end{equation}
for $k,i\in\mathcal{N}$ and $l\in\mathcal{L}$. By the definition
(\ref{eq:binary-transfer-variable}), if $d_{f,l}^{i}=1$, the CU
needs to send EN $i$ the subfile $(f,l)$ so as to enable multi-connectivity
transmission, since this subfile is not present in the cache of EN
$i$. We discuss some baseline uncoded fronthauling strategies and
then present the coded multicasting approach.

\subsection{Uncoded Unicasting\label{sub:Uncoded-Unicasting}}

With the baseline uncoded unicasting, the CU uses the fronthaul link
to send each EN $i$ the subfiles $(f,l)$ with $d_{f,l}^{i}=1$.
In this approach, the overlap between the sets of subfiles needed
by different ENs according to (\ref{eq:binary-transfer-variable})
is not taken into account. Therefore, the number $S_{B}$ of bits
transferred on the multicast link is given as
\begin{equation}
S_{B}=\sum_{i\in\mathcal{N}}\sum_{f\in\mathcal{F}_{\mathrm{req}}}\sum_{l\in\mathcal{L}}d_{f,l}^{i}\tilde{S},\label{eq:num-bits-uncoded-unicast}
\end{equation}
and the latency $T_{F}$ on the fronthaul link becomes
\begin{equation}
T_{F}=\frac{S_{B}}{C}.\label{eq:fronthaul-latency}
\end{equation}

\subsection{Uncoded Multcasting\label{sub:Uncoded-Multcasting}}

Since the multicast link is shared among the ENs, the subfiles needed
by multiple ENs need not be transferred separately to each EN. The
uncoded multicasting strategy hence transfers any subfile requested
by multiple ENs only once to minimize the number of fronthaul usages.
The number $S_{B}$ of bits multicast on the fronthaul link with this
approach is hence given as
\begin{equation}
S_{B}=\sum_{f\in\mathcal{F}_{\mathrm{req}}}\sum_{l\in\mathcal{L}}\mathsf{1}\left(\sum_{i\in\mathcal{N}}d_{f,l}^{i}>0\right)\tilde{S},\label{eq:num-bits-uncoded-multicast}
\end{equation}
where $\mathsf{1}(\cdot)$ denotes the indicator function, which returns
1 if the argument statement is true or 0 otherwise. Thus, the fronthaul
latency $T_{F}$ is given as (\ref{eq:fronthaul-latency}) with $S_{B}$
given as (\ref{eq:num-bits-uncoded-multicast}).

\subsection{Coded Multicasting\label{sub:Coded-Multicasting}}

Since the CU communicates with the ENs, each equipped with cached
contents, via a shared multicast link, coded multicasting can potentially
reduce the fronthaul overhead. Accordingly, the CU sends coded subfiles
obtained as linear combinations of the uncached subfiles in such a
way that the intended ENs can decode the coded subfiles based on the
cached subfiles (see, e.g., \cite{Ji-et-al}). The number $S_{B}$
of bits transferred on the fronthaul link in this approach is given
as $S_{B}=n_{\mathrm{sub}}\tilde{S}$, where $n_{\mathrm{sub}}$ is
the number of coded subfiles that are transferred on the fronthaul
link. This can be efficiently obtained by using the (suboptimal) greedy
constrained local coloring algorithm proposed in \cite[Sec. IV-A]{Ji-et-al}.
Having computed $S_{B}$ via the algorithm, whose details can be found
in \cite[Sec. IV-A]{Ji-et-al}, the fronthaul latency $T_{F}$ is
given as (\ref{eq:fronthaul-latency}).

As a final remark, we note that, for all the discussed fronthauling
strategies, the fronthaul latency $T_{F}$ increases with the connectivity
level $M$ due to the larger number of subfiles that need to be transferred
on the fronthaul link. This suggests that the optimal connectivity
level $M$ should be carefully selected by considering its conflicting
impacts on the edge and fronthaul latencies.

\section{Numerical Results\label{sec:Numerical-Results}}

In this section, we present some numerical results to compare the
latency of various fronthauling strategies discussed for the downlink
of an F-RAN system with a shared multicast fronthaul link. Throughout
the section, we set $\gamma=0.2$, $S=100$MB and $\alpha=0.7$. We
evaluate the total latency $T_{\mathrm{total}}$, the fronthaul latency
$T_{F}$ and the edge latency $T_{E}$ by averaging over the realizations
of the caching variables $\mathbf{c}$, the UEs' requests $\mathbf{f}$
and the channel matrices $\{\mathbf{H}_{k,i}\}_{k,i\in\mathcal{N}}$.

\begin{figure}
\centering\includegraphics[width=8.5cm,height=6.01cm]{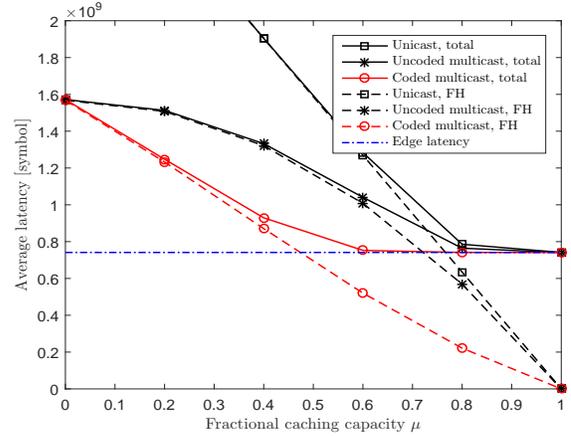}\caption{{\scriptsize{}\label{fig:graph-as-mu}Average latency $T_{\mathrm{total}}$
versus the fractional caching capacity $\mu$ for an F-RAN downlink
($F=60$, $L=50$, $N=4$, $n_{T}=n_{R}=1$, $C=2$, $P=20$ dB and
$M=2$).}}
\end{figure}

In Fig. \ref{fig:graph-as-mu}, we first investigate the impact of
the fractional caching capacity $\mu$ on the average latency $T_{\mathrm{total}}$,
the average fronthaul latency $T_{F}$ and the average edge latency
$T_{E}$ for an F-RAN downlink with $F=60$, $L=50$, $N=4$, $n_{T}=n_{R}=1$,
$\alpha=0.7$, $C=2$, $P=20$ dB and $M=2$. Note that the edge latency
$T_{E}$ is the same for unicast and multicast fronthaul transmissions.
It is observed that multicasting outperforms uncoded transmission,
particularly for small values of $\mu$, in which regime fronthaul
latency determines the overall latency. Furthermore, coded multicasting
can improve over uncoded multicasting, except for $\mu=0$, in which
case no side information is available at the ENs via caching, and
for $\mu=1$, in which case fronthaul transmission is unnecessary.
The gain is seen to be due to a reduction in the fronthaul latency.

\begin{figure}
\centering\includegraphics[width=8.5cm,height=6.01cm]{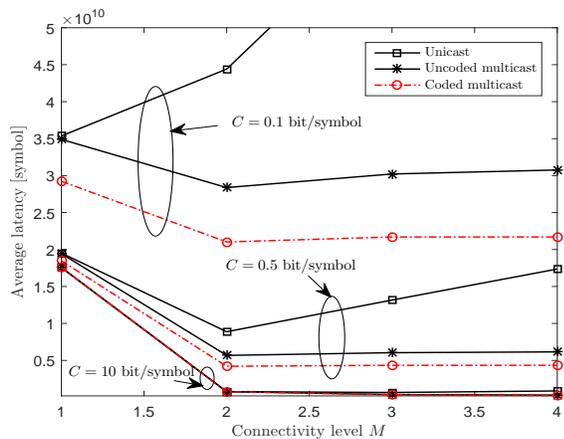}\caption{{\scriptsize{}\label{fig:graph-as-M}Average latency $T_{\mathrm{total}}$
versus the connectivity level $M$ for an F-RAN downlink ($F=50$,
$L=20$, $N=4$, $n_{T}=n_{R}=1$, $P=20$ dB and $\mu=0.3$).}}
\end{figure}

In Fig. \ref{fig:graph-as-M}, we plot the average latency $T_{\mathrm{total}}$
versus the connectivity level $M$ for an F-RAN downlink with $F=50$,
$L=20$, $N=4$, $n_{T}=n_{R}=1$, $P=20$ dB and $\mu=0.3$. The
figure confirms that there is an optimum connectivity level $M$ that
strikes the best trade-off between fronthaul and edge latencies. It
is also seen that the optimal value $M$ increases with the fronthaul
capacity $C$. Furthermore, the gain of coded multicasting is more
relevant for a larger connectivity level $M$ due to the increased
coding opportunities.

\begin{figure}
\centering\includegraphics[width=8.5cm,height=6.01cm]{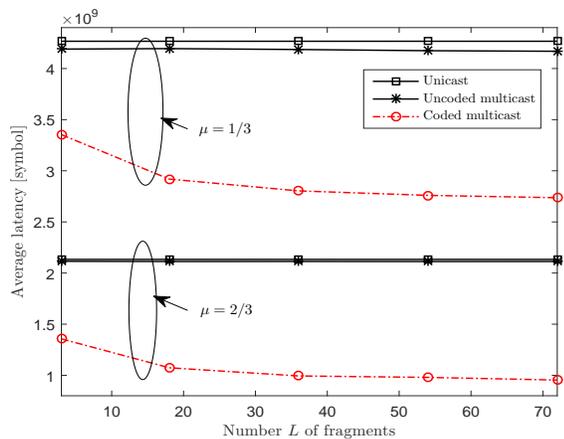}\caption{{\scriptsize{}\label{fig:graph-as-L}Average latency $T_{\mathrm{total}}$
versus the number $L$ of fragments for an F-RAN downlink ($F=60$,
$N=4$, $n_{T}=n_{R}=2$, $P=20$ dB, $C=0.5$ and $M=1$).}}
\end{figure}

A related conclusion can be reached from Fig. \ref{fig:graph-as-L},
which plots the average latency $T_{\mathrm{total}}$ versus the number
$L$ of fragments for an F-RAN downlink with $F=60$, $N=4$, $n_{T}=n_{R}=2$,
$P=20$ dB, $C=0.5$ and $M=1$. The figure shows that the gain of
coded multicasting becomes more significant for a larger $L$ due
to the larger number of coding opportunities.

\begin{figure}
\centering\includegraphics[width=8.5cm,height=6.01cm]{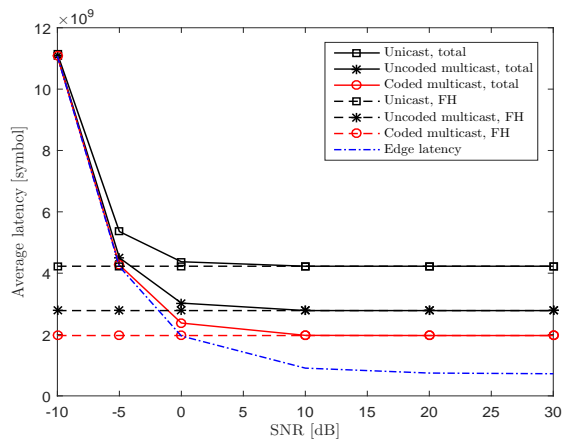}\caption{{\scriptsize{}\label{fig:graph-as-SNR}Average latency $T_{\mathrm{total}}$
versus the SNR $P$ for an F-RAN downlink ($F=60$, $L=60$, $n_{T}=n_{R}=1$,
$\mu=1/3$, $N=4$ $C=1.0$ and $M=2$).}}
\end{figure}

Lastly, in Fig. \ref{fig:graph-as-SNR}, we show the impact of the
SNR $P$ for an F-RAN downlink with $F=60$, $L=60$, $n_{T}=n_{R}=1$,
$\mu=1/3$, $C=1.0$, $N=4$ and $M=2$. The fronthaul latency does
not change with the SNR on the edge channel, and hence the latency
decrease is due to a reduction in the edge latency. As the SNR $P$
increases, the total latency $T_{\mathrm{total}}$ becomes limited
by the fronthaul latency while the edge latency dominates at lower
SNR values. As a result, the gain of coded multicasting is observed
in the regime of sufficiently large $P$.

\section{Conclusions\label{sec:Conclustion}}

This paper studies the delivery coding latency for the downlink of
an F-RAN system with a shared multicast fronthaul link. Under the
assumption of pipelined transmission on the fronthaul and edge links,
the advantages of coded multicast delivery on the fronthaul link were
investigated for multi-connectivity transmission across the ENs and
randomized fractional caching. We provided extensive numerical results
that validate the performance gains of the coded multicasting strategy
as compared to the conventional uncoded strategies. Among open problems,
we mention the development of an information-theoretic analysis that
accounts for the potential performance gains of coded multicast fronthauling
\cite{Koh-et-al}.

\end{document}